\documentclass[
preprint
,showpacs
,showkeys
,nofootinbib
,aps
,amsfonts
,amssymb
,byrevtex
,pdftex 
]{revtex4-1}

\pdfoutput=1
\usepackage{graphicx} 
\usepackage{amsmath, amssymb, bm} 
\usepackage[dvips]{color}

\usepackage{psfrag,epsfig}
\usepackage{natbib}
\usepackage{hyperref}
\usepackage{slashed}

\usepackage[normalem]{ulem}  

\def\){\right)} 
\def\({\left(} 
\def\]{\right]} 
\def\[{\left[}

\def\pnu{$p+p\rightarrow d+e^++\nu_e$}

\begin{document}


\title{Proton-proton fusion in lattice effective field theory}

\author{%
Gautam Rupak}
\email{grupak@u.washington.edu}

\author{%
Pranaam Ravi}
\email{pr340@msstate.edu}

\affiliation{Department of Physics $\&$ Astronomy and 
HPC$^2$ Center for Computational Sciences, 
Mississippi State
University, Mississippi State, MS 39762, U.S.A.}

\begin{abstract}
  The proton-proton fusion rate is calculated at low energy in a lattice effective field theory (EFT) formulation. The strong and the Coulomb interactions are treated non-perturbatively at leading order in the EFT. 
The lattice results are shown to accurately describe the low energy cross section within the validity of the theory at energies relevant to solar physics. In prior work in the literature, Coulomb effects were generally not included in non-perturbative lattice calculations. Work presented here is of general interest in nuclear lattice EFT calculations that involve Coulomb effects at low energy.  It complements recent developments of the adiabatic projection method for lattice calculations of nuclear reactions. 
\end{abstract}
\maketitle


\section{Introduction}
\label{sec_intro}

Calculations of nuclear reactions from a microscopic theory are of fundamental importance. Nuclear cross sections  are important in understanding the observed abundances of elements~\cite{Rolfs:1988,Burles:1999zt,Barwick:2004ep,Iocco:2008va,ThompsonNunes:2009}. These reactions occur under conditions of extreme densities and temperatures where all the known fundamental forces of nature -- gravitation, electro-weak interactions, and strong interactions -- play a role. Thus nuclear reaction cross sections impact disparate areas of physics such as astrophysics, nuclear physics and particle physics in a crucial manner.  The effective field theory (EFT) formulation of the microscopic nuclear interaction plays a central role in the nuclear reaction calculations~\cite{Bedaque:2002mn,Furnstahl:2008df,Epelbaum:2013tta,Machleidt:2011zz,Machleidt:2014hba}. EFT provides a model-independent framework  where one can make reliable estimates of the theoretical error. This is important as many of the nuclear reactions occur under extreme conditions that cannot be reproduced in terrestrial laboratories. Nuclear astrophysical models require reliable handle on the nuclear theory errors~\cite{Bahcall:1994,Burles:1999zt,Iocco:2008va}. Further, EFT provides a bridge between nuclear physics and particle physics where nuclear observables can be connected to particle physics parameters such as the quark masses~\cite{Epelbaum:2012iu}. 

Applications of EFT in the few-nucleon systems have been quite successful~\cite{Bedaque:2002mn,Furnstahl:2008df,Epelbaum:2013tta,Machleidt:2011zz,Machleidt:2014hba}. Though there is a good understanding of the microscopic nuclear interactions, its application to larger nuclear systems poses serious computational challenges. Numerical lattice methods from particle physics combined with EFT provide a promising possibility. The lattice EFT formulation  allow a systematic error analysis derived from  EFT.  Ground and excited state energies for several atomic nuclei have been calculated accurately~\cite{Rwall,Epelbaum:2010xt,Epelbaum:2012qn}. Many-body properties in  dilute neutron matter have also been addressed~\cite{Borasoy:2007vk}.  Recently progress has been made in calculating nuclear reactions using lattice methods albeit in simple systems~\cite{Rupak:2013aue,Pine:2013zja,Elhatisari:2014lka}. The proposal in Refs.~\cite{Rupak:2013aue,Pine:2013zja} is to first construct an effective two-body Hamiltonian from first principle using an adiabatic  projection method. This Hamiltonian is then used to calculate elastic and inelastic reactions involving nuclei such as $a+b\rightarrow \gamma+c$, $a+b\rightarrow c+d$ with $a$, $b$, $c$ being atomic nuclei and $\gamma$ a photon. 
In this work  we consider the contribution from the long range Coulomb force. Nuclear reactions involving compound nuclei will necessarily involve Coulomb interactions that become non-perturbative at energies relevant to astrophysics. To test the basic formulation we calculate proton-proton elastic scattering and fusion at low energy. This simpler system allows us to isolate the Coulomb effect without a complicated nuclear strong force.

The pioneering calculation by Bethe and Critchfield showed that  proton-proton fusion  \pnu ~powers the sun~\cite{BetheCritchfield:1938,Bethe:1939}.
This is a rare weak process that is the first crucial step in solar fusion.  A small Coulomb barrier along with the slow rate of the weak process leads to a long and steady burning of hydrogen in medium mass stars such as our sun~\cite{bertulaniBook}.  The proton fusion rate is crucial to understanding  solar neutrino production and its subsequent detection in terrestrial laboratories~\cite{Adelberger:1998qm}.

Bahcall and May refined the fusion rate calculation~\cite{BahcallMay:1969} and set the benchmark for future evaluations 
such as Refs.~\cite{Bahcall:1994,Schiavilla:1998}.  
The capture rate was expressed in terms of model-independent parameters such as the deuteron binding momentum, the proton-proton scattering length, etc.,  that are  not sensitive to the details of the nuclear potential. The proton-proton fusion rate was analyzed in EFT with short-ranged interactions in Refs.~\cite{Kong:1999tw,Kong:1999mp}. The EFT calculations matched the work by Bahcall and May when expressed in terms of the two-body scattering parameters and one-body currents. Higher order corrections from two-body currents have also been included in EFT calculations in a systematic manner, see Ref.~\cite{Butler:2001jj}.
We consider the leading order (LO) contribution in lattice EFT.   Both the strong and Coulomb interaction are non-perturbative at LO. The higher order contributions are perturbative~\cite{Kong:1999tw,Kong:1999mp,Butler:2001jj} and should not pose any technical challenge in future lattice calculations.

\section{Interaction}
\label{sec_interaction}
Proton-proton fusion at solar energies around the Gamow peak is dominated by capture from the $s$-wave. At these energies $E\sim 6$ keV, the initial state proton-proton interaction at LO is described by the Lagrangian~\cite{Kong:1998sx,Kong:1999sf}: 
\begin{align}\label{eq:StrongL}
\mathcal L=\psi^\dagger\[i\partial_0+\frac{\nabla^2}{2 M}\]\psi-\frac{c_0}{4}(\psi\sigma_2\psi)^\dagger(\psi\sigma_2\psi),
\end{align}  
where the proton mass $M=938.3$ MeV, and $\psi$ represents the spin-1/2 protons. The Pauli matrix $\sigma_2$ is used to project the protons onto the spin-singlet channel. We use natural units with $\hbar=1=c$. The coupling $c_0$ can be determined from proton-proton scattering length $a_p$~\cite{Kong:1998sx,Kong:1999sf}. The strong interaction potential in coordinate space  for proton-proton scattering in the $s$-wave spin-singlet channel, corresponding to Eq.~(\ref{eq:StrongL}), is 
\begin{align}
V_s(\vec{\bm{r}})=c_0\delta(\vec{\bm{r}}). 
\end{align}
 The long range Coulomb force is described by the Coulomb potential
 \begin{align}
 V_c(\vec{\bm{r}})=\frac{\alpha}{r}, 
 \end{align} 
 with $\alpha=1/137$. 
 Given these interactions, we construct the lattice theory by discretizing space in a periodic box. The strong interaction potential reduces to a Kronecker delta function at the origin on the lattice. The Coulomb potential is defined on the discretized  lattice in a straightforward manner. However, at the origin we regulate it, i.e., replace it by a Kronecker delta function with a coupling $d_0$ to be determined later.   In the presence of both strong and Coulomb potentials, only the linear combination of $c_0+d_0$ determines phase shifts and amplitudes. This is a consequence of the overlap of the ultraviolet divergences in the strong and Coulomb interactions in the EFT~\cite{Kong:1998sx,Kong:1999sf}.

Proton-proton fusion involves a deuteron in the final state that can be described in the EFT accurately~\cite{Chen:1999tn}. The LO spin-triplet interaction can be described with a short-ranged interaction $(\psi\sigma_2\sigma_i\phi)^\dagger (\psi\sigma_2\sigma_i\phi)$ where $\phi$ is the spin-1/2 neutron field. The coupling for this spin-triplet interaction is tuned independently of the spin-singlet interaction in Eq.~(\ref{eq:StrongL}) to reproduce the deuteron binding energy $B=2.2246$ MeV~\cite{Leun:1982}. The deuteron bound state can be described in the lattice formulation of the short-ranged interaction as well. 
 
 \section{scattering and fusion}
 \label{sec_scattering}
 Elastic scattering is commonly described in lattice calculations using L\"uscher's method~\cite{Luscher:1986pf,Luscher:1991ux}.  The energy shifts in a periodic box in the presence of a short-ranged interaction is used to calculate the elastic phase shifts. Perturbative Coulomb contributions to two-particle scattering in a finite volume have been considered recently~\cite{Beane:2014qha} but 
 a general method for calculating Coulomb interactions non-perturbatively at 
low energy using  L\"uscher's method doesn't exist. Here we calculate the non-relativistic phase shift in the presence of the long range Coulomb force using a hard spherical wall boundary condition.  This method was introduced in Ref.~\cite{Rwall} to calculate two-body phase shift due to the short-ranged strong interaction. The spherical wall method was found to be better suited than L\"uscher's method for problems involving coupled channels~\cite{Rwall}.

To understand the spherical wall method, consider a short-ranged potential $V_s(r)$ inside a hard spherical wall of radius $R$~\cite{Rwall}. The continuum asymptotic $s$-wave solution to the Schr\"odinger equation inside the hard wall has the form 
\begin{align}\label{eq:psiStrong}
\left|\psi_s(\vec{\bm r})\right|=\left| j_0(kr)\cos\delta_s -y_0(k r)\sin\delta_s\right|,
\end{align}
in terms of the spherical Bessel functions $j_0$ and $y_0$ at the center-of-mass momentum $k$. At the spherical wall boundary $R$, the wave function must vanish, giving 
 \begin{align}\label{eq:phaseStrong}
 \tan\delta_s(k)=\frac{j_0(kR)}{y_0(kR)}. 
 \end{align}
On a cubic lattice  one cannot fit a sphere of radius $R$. Instead for a given spherical hard wall radius $R$, one defines an adjustable wall radius $R_w$ where the free wave function vanishes~\cite{Rwall}:
 \begin{align}\label{eq:Rwall}
 j_0(k_0 R_w)=0\Rightarrow R_w=\frac{\pi}{k_0}.
 \end{align} 
 $k_0$ is the  center-of-mass momentum of the free theory on the lattice. It corresponds to the momentum of the first energy of the spectrum on the lattice. The self-consistent use of $R_w$ in Eq.~(\ref{eq:phaseStrong})  by setting $R=R_w$ is shown to accurately reproduce the strong interaction phase shifts for various two-nucleon channels~\cite{Rwall}. We follow the same procedure in calculating the  strong-Coulomb phase shift for proton-proton scattering on the lattice.

 Traditionally, proton-proton scattering is described by considering the Coulomb subtracted phase shift $\delta_{sc}=\delta_\mathrm{full}-\delta_c$. The full phase shift specifies the scattering amplitude $\mathcal T$ through the relation
\begin{align}
T(k)=\frac{2\pi}{\mu}\frac{\exp(i2\delta_\mathrm{full})-1}{2i k},
\end{align}
where $\mu=M/2$ is the reduced mass.  The purely Coulomb phase shift $\delta_c(k)=\operatorname{Arg}[\Gamma(1+i\eta_k)]$ is independent of the short-ranged nuclear interaction.  A standard result derived from the analytical properties of the scattering amplitude \cite{Bethe:1949,Kok:1982} gives:
\begin{align}\label{eq:phaseSCanalytic}
C_{\eta_k}^2\left[ k\cot\delta_{sc}-i k\right]+2k\eta_k H(\eta_k)=-\frac{1}{a_p}+\frac{r}{2}p^2 +\mathcal O(p^4). 
\end{align}
See Ref.~\cite{Koenig:2012bv} for a recent derivation. 
The Sommerfeld factor 
\begin{align}
C_{\eta_k}^2=\frac{2\pi\eta_k}{\exp(2\pi\eta_k)-1},
\end{align}
represents the probability of finding two protons at the origin. $\eta_k=\alpha\mu/k$ is the Sommerfeld parameter that serves as the formal expansion parameter in Coulomb interactions. A small $\eta_k$ implies Coulomb effect is perturbative. At low momentum $k$, $\eta_k$ is large and Coulomb effect has to be treated non-perturbatively. 
The function $H(\eta_k)$ is defined through the di-gamma function $\psi(x)=\partial_x\ln\Gamma(x)$ as
\begin{align}
H(\eta_k)=\psi(i\eta_k)+\frac{1}{2i\eta_k}-\ln(i\eta_k). 
\end{align}
In the limit $\alpha\rightarrow 0$ for ``neutral" protons, $\eta_k H(\eta_k)=i/2$ and Eq.~(\ref{eq:phaseSCanalytic}) reduces to the familiar effective range expansion for short-ranged interaction 
\begin{align}
k\cot\delta_{sc}=-\frac{1}{a_p}+\frac{r}{2}p^2 +\mathcal O(p^4).
\end{align}
For the Coulomb calculation we use Eq.~(\ref{eq:phaseSCanalytic}) with the experimentally determined $a_p=-7.82$ fm and set $r=0$ for our LO calculation. In effect we use one experimental input $a_p$ to determine a single coupling $\hat{c}_0+\hat{d}_0$ (in lattice units).  Next-to-leading order (NLO) correction related to the effective range can be added to the EFT calculation systematically~\cite{Kong:1999sf}.

The lattice calculation of the Coulomb subtracted phase shift $\delta_{sc}$ follows from the conventional definition of the $s$-wave proton-proton wave function in the presence of both the strong and the Coulomb force 
\begin{align}\label{eq:psiSC}
\left|\psi_p(\vec{\bm r})\right| = \frac{\left|F_0(kr)+G_0(k r)\tan\delta_{sc}\right|}{k r},
\end{align}
that resembles the solution in Eq.~(\ref{eq:psiStrong}) with $F_0(kr)$ and $G_0(kr)$ being the regular and irregular Coulomb wave functions, respectively. See Ch. 14 in Ref.~\cite{Abramowitz}. 
Requiring the wave function vanish at the spherical wall gives the lattice phase shift as
\begin{align}
\delta^{(\mathrm{latt})}_{sc}(k)=\tan^{-1}\[-\frac{F_0(k R_w)}{G_0(kR_w)}\],
\end{align}
where $R_w$ is again determined from the free particle spectrum in Eq.~(\ref{eq:Rwall}).  

\begin{figure}[thb]
\begin{center}
\includegraphics[width=0.6\textwidth,clip=true]{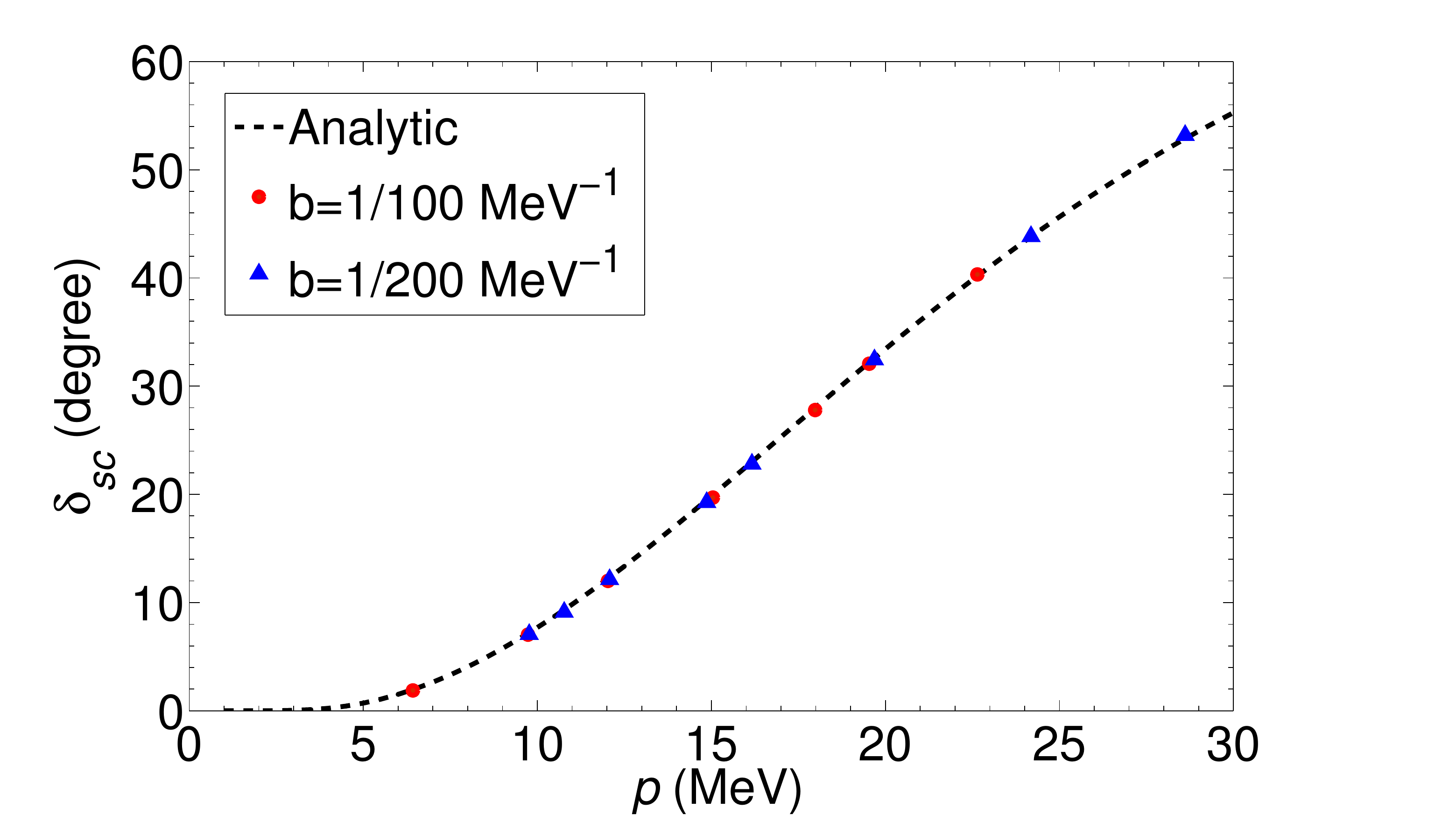}
\caption{\protect Coulomb subtracted phase shift. The dashed curve is the analytical result from Eq.~(\ref{eq:phaseSCanalytic}). The red circle data correspond to $b=1/100$ MeV$^{-1}$, $\hat{c}_0+\hat{d}_0=-0.4047$, and the blue triangle data correspond to $b=1/200$ MeV$^{-1}$, $\hat{c}_0+\hat{d}_0=-0.8330$ lattice results, respectively. }
\label{fig:phaseSC}
\end{center}
\end{figure}

The results of the lattice calculations are shown in Fig.~\ref{fig:phaseSC}. We reproduce the analytical result at two different lattice spacings $b=1/100$ MeV$^{-1}$ and $b=1/200$ MeV$^{-1}$. The coupling $\hat{c}_0+\hat{d}_0$ depends on the lattice spacing. This is expected since there is a short-distance scale associated with the dimensionful  coupling $c_0$ in the continuum theory, Eq.~(\ref{eq:StrongL}),  that is regulated by the lattice spacing. This is reflected in the scale dependence of the linear combination  $\hat{c}_0+\hat{d}_0$.  We find that the lattice calculations reproduce Coulomb subtracted  $\delta_{sc}$ accurately to within about 3\% or less. 

After having calculated the elastic scattering phase shift, we consider the proton-proton fusion rate. At low energy the fusion rate is dominated by peripheral collision.  The nuclear contribution is described by the expectation value of $\langle\psi_d|\operatorname{O}_{EW}|\psi_p\rangle$, where $\psi_d$ is the final state deuteron wave function, $\psi_p$ is the incoming proton-proton wave function, and $\operatorname{O}_{EW}$ is the electro-weak current. The strong interaction contribution to the transition amplitude from one-body current at LO can be written as~\cite{BahcallMay:1969}
\begin{align}
\mathcal T_\mathrm{fusion}(k)=\int d^3r\ \psi_d^\ast(\vec{\bm r})\psi_p(\vec{\bm r}),
\end{align}
with the deuteron bound state wave function
\begin{align}
\psi_d(\vec{\bm r})=\sqrt{\frac{\gamma}{2\pi}}\frac{\exp(-\gamma r)}{r},
\end{align} 
where the binding momentum $\gamma=\sqrt{2\mu B}$ is defined in terms of the deuteron binding energy $B=2.2246$ MeV.  The incoming $s$-wave proton-proton  
wave function $\psi_p(\vec{\bm r})$ is given by Eq.~(\ref{eq:psiSC}).  It is possible to evaluate the deuteron wave function using the lattice formulation. Here we want to focus on the non-perturbative Coulomb calculation as bound state wave functions have been calculated accurately in lattice EFT before~\cite{Rwall,Epelbaum:2010xt,Epelbaum:2012qn}. 
It is convention to consider the normalized matrix element following Salpeter~\cite{Salpeter:1952}
\begin{align}
\Lambda(k)=\sqrt{\frac{\gamma^3}{8\pi C_{\eta_k}^2}} \left| \mathcal T_\mathrm{fusion}(k)\right|. 
\end{align}

\begin{figure}[thb]
\begin{center}
\includegraphics[width=0.6\textwidth,clip=true]{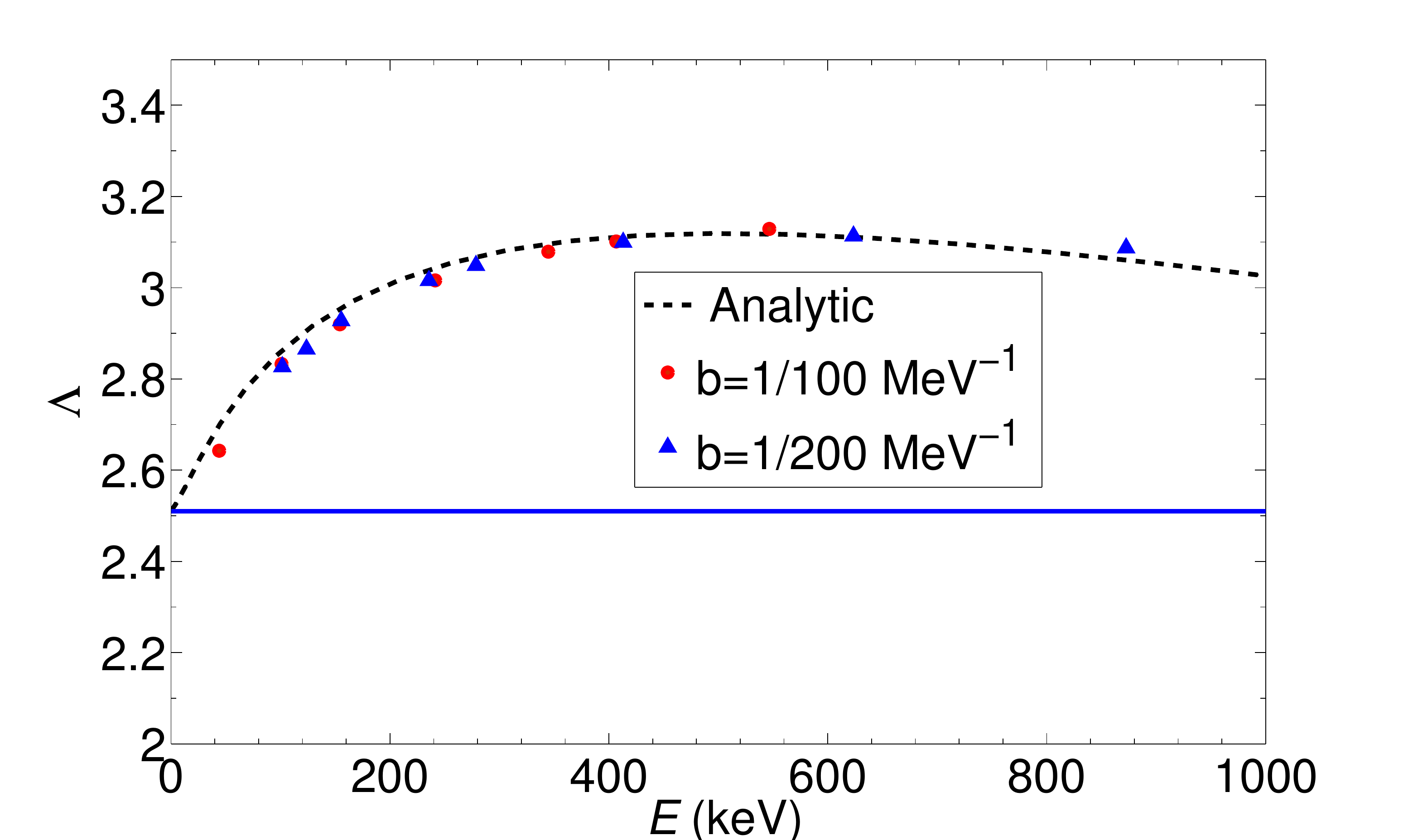}
\caption{\protect Proton-proton fusion rate. The dashed curve is the  analytical result. Red circles and blue triangles show lattice data at $b=1/100$ MeV$^{-1}$ and $b=1/200$ MeV$^{-1}$ with the lattice couplings from Fig.~\ref{fig:phaseSC}. The  horizontal grid line at $\Lambda=2.51$ indicates the LO continuum EFT result at zero energy~\cite{Kong:1999tw,Kong:1999mp}.}
\label{fig:fusion}
\end{center}
\end{figure}

We use the lattice values for $\delta^{(\mathrm{latt})}_{sc}$ to calculate $\Lambda(k)$. The results are compared with the analytical result in Fig.~\ref{fig:fusion}. Given that the Coulomb subtracted phase shifts are calculated accurately on the lattice, see Fig.~\ref{fig:phaseSC}, it is expected that the lattice results for $\Lambda(k)$ would agree well with the analytical results. The lattice results agree with the analytical results to about 3\% or less. 
An extrapolation to zero energy gives $\Lambda^{(\mathrm{latt})}(0)=2.49\pm0.02$ which is consistent with the LO continuum EFT calculation $\Lambda^{(\mathrm{EFT})}=2.51$~\cite{Kong:1999tw,Kong:1999mp}. At solar energies $E=k^2/(2\mu)\sim 6$ keV, $\Lambda(E)$ deviates no more than a few percent from $\Lambda(0)$. Beyond LO, effective range corrections in both the incoming proton-proton channel and the bound state deuteron channel contributes~\cite{Kong:1999tw,Kong:1999mp}. Contribution from mixing between the $s$-wave and the $d$-wave component of the deuteron can be included as well~\cite{BahcallMay:1969,Butler:2001jj}. Two-body currents contribute as well at higher order~\cite{Butler:2001jj}.

\section{Conclusions}
\label{sec_conclusions}
We calculated the elastic proton-proton phase shift on the lattice for momenta $p\lesssim 30$ MeV. The lattice EFT was described using a short-ranged nuclear interaction at LO, in addition to the Coulomb force. While  L\"uscher's method is not applicable to problems involving the long-ranged non-perturbative Coulomb force, the spherical wall method was found to be an effective
approach for the calculation.  The LO coupling was determined from the known scattering length parameter. The lattice results were shown to agree with the known analytical results to about 3\% or less.

The strong interaction contribution to the proton-proton fusion rate in the presence of the Coulomb force was calculated. The fusion rate is proportional to the bound state deuteron wave function $\psi_d(\vec{\bm r})$ and the incoming $s$-wave proton-proton wave function $\psi_p(\vec{\bm r})$. The wave function 
$\psi_p(\vec{\bm r})$ is determined by the Coulomb subtracted phase shift $\delta_{sc}$ that is calculated accurately on the lattice. 
The lattice fusion rate calculations reproduced the continuum results. Future work should explore the higher order corrections in the lattice EFT calculation.  Contributions from effective range,  $s$-$d$ mixing in the deuteron wave function, two-body electro-weak currents at higher order in the continuum EFT are well known from the work in Refs.~\cite{Kong:1999tw,Kong:1999mp,BahcallMay:1969,Butler:2001jj}. Lattice calculation of these higher order effects should be explored. 

The results presented in this work complement the recent work in the adiabatic projection method for calculating nuclear reactions on the lattice. The adiabatic projection method allows the construction of an effective two-body Hamiltonian from first principle to describe certain low energy electro-weak  nuclear reactions. Nuclear reactions involving compound nuclei are of importance in astro, nuclear and particle physics analysis. The Coulomb force is expected to play an important role in these low energy reaction calculations. The current work would be part of the program to study these reactions using lattice EFT.

\begin{acknowledgments}
The authors thank Dean Lee for many valuable discussions and comments. Renato Higa is thanked for useful comments on the manuscript.   
Partial support was provided by the U.S. National Science Foundation  grant No. PHY-1307453.  Computational resources was provided by the HPCC at MSU. 
\end{acknowledgments}

%

\end{document}